\documentclass[12pt,preprint]{aastex}
\usepackage{psfig}
\advance \voffset by 1.00cm\relax

\begin{document}

\title{Population synthesis of neutron stars, strange (quark) stars and
black holes}

 \author{Krzysztof Belczynski\altaffilmark{1,2,4}, 
         Tomasz Bulik\altaffilmark{2} and 
         W{\l}odzimierz Klu{\'z}niak\altaffilmark{3}}

 \affil{ 
     $^{1}$ Northwestern University, Dept. of Physics
     \& Astronomy, 2145 Sheridan Rd., \#F325, Evanston, IL 60208, USA;\\
     $^{2}$ Nicolaus Copernicus Astronomical Center,
     Bartycka 18, 00-716 Warszawa, Poland\\
     $^{3}$ Institute of Astronomy, ul. Lubuska 2, 
     Zielona G\'ora University,
     65-265 Zielona G\'ora, Poland\\
     $^{4}$ Lindheimer Fellow 
}      
 
 \begin{abstract} 

We compute and present the distribution in mass of single and
binary neutron stars, strange stars, and black holes. The
calculations were performed using a stellar population synthesis
code. 
We follow evolution of massive single stars as well as binaries with 
high mass primaries.
The final product of the latter evolution can be either a binary
composed of a white dwarf and a compact object (neutron star,
black hole, strange star), two compact objects in a binary, or
two single stars if the system was disrupted. We find in
binaries a population of black holes which are more massive than
single black holes which are a product of either binary or
single evolution. We also find that if quark stars exist at all,
their population can be as large as the population of black
holes.

 \end{abstract}


\section{INTRODUCTION}

Binary population synthesis is a useful tool for studying  the
statistical properties of stars, including the compact objects
(e.g., Pols \& Marinus 1994; Bethe \& Brown 1998;
Portegies-Zwart \& Yungelson 1998;  Bloom, Sigurdsson, \& Pols
1999; Belczynski \& Bulik 1999). Compact objects are stellar
remnants of size much smaller than that of white dwarfs,
so according to current views they could
be either black holes or neutron stars or, possibly, quark stars. 

We wish to address the
following questions: What is the distribution of masses of the
compact objects formed along different  evolutionary paths?  Given
the distribution of compact object masses, what are the relative
numbers of different types of objects (neutron stars, quark
stars, black holes) both single and in binaries? What fraction
of binaries give rise to single compact objects, and what fraction
survives as binaries and of what type?

In Section 2 we shortly describe the population synthesis code
used here, in Section 3 we summarize what is known about
the masses of neutron stars, quark stars and
of black holes. In Section 4 we discuss the constraints on the
masses of compact objects, in Section 5
we present the results, and finally in
Section 6 we give the conclusions.

\section{POPULATION SYNTHESIS CODE}

We use {\em StarTrack}, a stellar binary population synthesis code  
consisting of two parts. The single star evolution is based on the
formulae from Hurley et al. (2000), modified as follows.
  We have changed the
prescription for mass of the compact object formed in a
supernova explosion. We use the original Hurley et al. (2000)
formulae to obtain final CO  core mass. We use models of Woosley
(1986) to calculate the final  FeNi core mass (for a given CO
core mass), which will collapse and form a compact  object
during supernova explosion. Finally, we include calculations of
Fryer and Kalogera (2001) to  take into account black hole
formation both through direct collapse and partial fall back.

The binary evolution is described in Belczynski, Kalogera \&
Bulik (2002) and Belczynski 2001.  
We evolve only binaries where at least one star  will 
undergo  a supernova explosion and form a compact object.  
The evolution starts at zero-age main sequence.
During the course of evolution  we include the following effects
as appropriate: wind mass loss (standard, Wolf-Rayet, luminous blue variables),
tidal circularization of binary orbit, conservative/nonconservative 
mass transfer, common envelope evolution, rejuvenation, hyper-accretion 
onto compact objects, detailed supernova explosion treatment.

Many binaries are disrupted in supernova explosions, as a result
of mass loss and the natal kick. For supernova kicks we use the 
distribution presented by Cordes \& Chernoff (1998).  We use
smaller kicks when the compact object is a black  hole formed
via partial fall-back, and no kicks for the black holes formed through 
direct collapse, for details see Belczynski et al. (2002). 
We continue to evolve each star,  until the  formation of a 
stellar remnant.  At the endpoint of binary evolution either two
single remnants are left or a binary - in either case at least
one of the remnants is a compact object, while the other is
either a compact object or a white dwarf.

\section{QUARK STARS vs. NEUTRON STARS}

Bodmer (1971) suggested that stars composed of up, down and strange quarks
(in roughly equal numbers) may exist if quark plasma is the ground state
of matter. Relativistic models of ``strange'' stars composed of such
self-bound quark matter were first computed
by Brecher \& Caporaso (1976), Witten (1984),
Alcock, Farhi \& Olinto (1986),
 and  Haensel, Zdunik \& Schaeffer (1986). Alcock et al. (1986)
give a detailed discussion of the possible avenues of formation of quark
stars. If they are formed through a phase transition after a certain
critical density is exceeded in the core of a neutron star, or
if they are formed in a supernova of a star which has captured
a ``seed'' of strange matter, quark stars could be more massive
than neutron stars. In other scenarios, no neutron stars at all
would exist, or the abundance of quark stars need not be a function
of their mass.
The astrophysics of quark stars has recently been reviewed by
Cheng, Dai and Lu (1998) and Madsen (1999). 

 It has been
argued that young, glitching, pulsars cannot be strange stars
 (Alpar 1987).  Madsen (1988) and Caldwell \& Friedman (1991)
 argue that strange stars in Hulse-Taylor type
binaries would eventually contaminate the entire Galaxy with strange matter
as a result of  their binary coalescence,
and thus preclude the formation of young neutron stars.
But Klu\'zniak (1994) pointed out
that many millisecond pulsars could in principle be strange stars,
and Cheng \& Dai (1996) suggest
that strange stars could be formed through accretion induced
phase transition in low-mass X-ray binaries (LMXBs).
The compatibility of strange star models with
 the observed kHz QPOs frequencies in LMXBs was discussed by
 Bulik, Gondek-Rosi\'nska \& Klu\'zniak (1999), Zdunik et al. (2000)
and others.
Simulations of the hydrodynamics of
coalescence indicate that quark matter is not always expelled in the binary
merger of a strange star with another compact object
(Lee, Klu\'zniak, \& Nix 2001), so co-existence of neutron stars
and strange stars may be possible, after all.

The existence of self-bound quark matter remains a hypothesis, and we
cannot give definitive conclusions as to the abundance of quark stars.
However, if one assumes that, say, quark stars constitute
a fraction $f$ of compact objects
in some mass range  $(M_1, M_2)$ (and possibly a different fraction in
some other mass range), then the number of quark stars can be read off
from our plots of differential distributions. We will find, that if
$f$ is a sizable fraction of unity, and if $(M_1, M_2)$ is not too narrow,
then the number of quark stars may be comparable to the number of
black holes.

\section{MASSES OF COMPACT OBJECTS}

During the evolution we make nearly no distinction between the various
types of compact objects, they only differ in mass
 (and in the natal kick distribution).
In presenting the results, we assume that a compact object
exceeding  a certain mass is
always a black hole.
The actual value of the maximum mass of a neutron star is not 
known---the value
of mass above which a stable neutron star can no longer exist
depends  on the unknown equation of state  of matter at
supernuclear densities. Models (Arnett \& Bowers 1976,
Kalogera and Baym 1996) give  maximum masses 
from $1.4\,M_\odot$ to above $2.9 \, M_\odot$.
For the maximum masses of moderately rotating quark stars
in the MIT bag model as a function of the unknown bag constant
 see Zdunik et al. (2000),
and for the masses of quark stars when the
Dey et al. (1998) equation of state is used, see
Gondek-Rosi{\'n}ska et al. (2000).
 The maximum mass
depends also on the rotation of the neutron star
(Friedman, Parker \& Ipser 1986; Cook, Shapiro \& Teukolsky 1992).
The corresponding increase of maximum mass in rapidly
rotating quark stars is even larger (Stergioulas, Klu\'zniak \& Bulik 1999).
However, in this
discussion we neglect these effects of stellar rotation, 
i.e., we assume none of
 the neutron stars (or quark stars) formed
has a period less than $\sim 10$ms.

Of course, there is no theoretical maximum to the mass of a black
hole. The maximum masses we find in our calculations simply reflect the
formation route of the black hole. We find that the most massive
black holes survive in binaries (Fig.~\ref{fig1}, compare panels labeled ``group 0''
or ``group 1'' with the ones labeled ``group 2'' or ``group 3'').

Observations of binary stars yield direct information on masses
of some compact objects. Neutron stars in the Hulse-Taylor type
binaries have accurately measured masses of $1.44$ and
$1.39$\,M$_\odot$. Millisecond
pulsars have been analyzed by Thorsett \& Chakrabarty (1999)
who found that they are consistent with all being in the narrow
mass range of $1.34 \pm 0.04 \,M_\odot$  Among the neutron stars
which exhibit X-ray bursts the mass of Cyg X-2 has been quoted
as $1.78 \pm 0.23\, M_\odot$  (Orosz \& Kuulkers 1999).
However, for most low mass x-ray binaries (LMXBs) the masses
remain unknown.

There is a class of LMXBs where bright X-ray emission is
transient and the masses of the compact object cluster in the
range $\sim 5.5$ to $\sim 7.5 M_\odot$. These are thought to be
black holes. At present our code does not yield an excess  of
black holes in this mass
range with a white dwarf companion in the binary, 
instead a peak at about $10M_\odot$ results (see Fig.~\ref{fig1}).
However, we note that according to our results, single black
holes are particularly abundant at $M\sim 5 M_\odot$, and there is
a deficit of single compact objects in the mass range of about
$2.5M_\odot$ to $5M_\odot$. If the black hole LMXBs were formed
through binary capture in globular clusters, our results would be
consistent with the measured masses of the transient sources.

\section{RESULTS}

Compact objects may be formed both through  single and binary
stellar evolution. Single compact objects may be  descendants of
massive single stars but also  of components of a binary
system  disrupted in a supernova explosion.  We will denote the
single compact objects formed from primordial single stars as 
group 0, and formed as a result of the binary evolution as 
group 1.  Under favorable conditions some binaries survive
supernova explosions, and they finally form tight systems with
compact object/objects. Most of these binaries will consist of a
white dwarf and a compact object and the rest will form binaries
with two compact objects  (we will denote the compact object in
binaries with white dwarfs as group 2, and the double compact
objects as group 3).

Figure~\ref{fig1} shows the number of compact objects  per mass interval 
formed along each route. We start forming compact objects at
mass $\sim 1.2 M_\odot$ and their number 
falls off with the mass of the final compact object, as expected
for our assumed  initial mass function $\sim M^{-2.7}$. The peak
in the distribution in Figure~\ref{fig1}  around $\sim 10 M_\odot$
reflects the relation we obtain  between ZAMS mass of a
progenitor and the final mass of a compact object. This relation
for a wide range of progenitor ZAMS masses results in a final
compact object mass of $\sim 10 M_\odot$ (Belczynski et al. 2002). 
This is an effect of stellar wind which increases with the
 mass of the star, and  thus decreases  the final mass of a compact
object for large initial stellar masses. As a result 
the mass of the FeNi core is a weak function of the initial 
mass of the star  for a wide range of 
the ZAMS masses. 

In Figure~\ref{fig2} we present, for each 
formation route separately,
  the cumulative fraction of compact
objects as a function of their final mass.  
The normalization is such, that a fraction of unity corresponds to  the total
number of stars used in the simulation (we used a total of
$7\times 10^6$ binaries and $7\times 10^6$ single stars), and assumed a binary 
fraction of 50\%, i.e., we assumed that (initially) out of every three stars
one is single and the other two are in a binary system.
 The  single star and the primary mass in binaries was in
the range  $5$ to $100 M_\odot$, and the mass of the secondary was
found assuming a flat mass ratio $q$ distribution and
a $-2.7$ slope of the initial mass function. Each
distribution rises quickly in the small mass range, which is also
seen in Figure~\ref{fig1}. Thus within each group  even a small mass
window $M_2-M_1$ may yield a significant number of quark stars,
if they constitute a sizable fraction of the objects in that mass window.
For example, if quark stars are formed in the narrow mass range $(M_1,M_2)$,
with $M_1=1.7 M_\odot$ and $M_2=1.8 M_\odot$, and no neutron stars of
that mass exist, the fraction
of quark stars in each group will be from a few to ten percent. 
This fraction is comparable  to that of black holes in any given
group, which is  about $15-20$\% in groups 0, 1, and 2, and 
$\approx 50$\% in group 3.  We also note that  most of strange
stars in the Galaxy   should exist as single objects, and only a
small fraction  of them $\approx 10^{-5} - 10^{-4}$ is going to
be in double compact object binaries. 

We have listed in Table~\ref{tab1} the numbers of binaries with compact object
components obtained in our simulation. 
The binaries are classified according to their component masses, and for 
illustrative purpose we have labeled the objects in the mass range
$1.7 M_\odot<M<2.5 M_\odot$ as strange stars.
The table allows to read the relative numbers of objects of
different type.

\section{DISCUSSION}

We have shown the effects of the binary evolution on the
distribution of masses of compact objects. As expected the bulk
of the population of compact objects have masses below
$2\,M_\odot$. While for single stellar evolution  there exists a
unique relation between the stellar mass and  the compact object
mass,  there is no such relation when the binary evolution is
taken into account. Binary evolution works both ways, the mass
of a compact object formed from a particular star  in a binary 
can be  smaller or larger than that formed from an identical
star undergoing  single stellar evolution.

In the low
mass range, the cumulative fraction of compact objects
rises steeply with increasing mass--see Figure~\ref{fig2}.
 Thus, even in a small mass interval
($M_1,M_2$),  the fraction of stars in the compact
object population can be large. On the
other hand the fraction of black holes hardly depends on their minimum
mass
(the cumulative curves flatten above $3\,M_\odot$). 
We conclude
that the population of quark stars can easily be as large as the
population of black holes, even if there is only a small mass
window for their formation.

The low mass peak in
differential distribution of Figure~\ref{fig1}, is less pronounced for
double compact object binaries, group 3.  Thus a chance of
finding  quark stars in Hulse Taylor type objects is slim
primarily because of the small number of
such objects known so far. The prospects look better for compact
objects in binaries with dwarfs. Although Thorsett \& Chakrabaty
(1999) show that  the masses of these objects are consistent
with being constrained to a narrow range, the mass function for
individual objects allows different (higher or lower) masses.
However the observed number of these sources is not large
and the search here may suffer  from small number statistics.
Our  results show that most quark stars should exist as single
objects, yet it is the most difficult to measure the
masses and radii for them. Therefore the search for quark stars
may have to  concentrate on single compact objects, such as
pulsars.

It is interesting to note that the most massive black holes
survive in binaries. This is related both to the difficulty of
disrupting a binary with a very massive black hole 
and to the need for fallback in forming such massive black holes
with $M>10M_\odot$.

\acknowledgements
KB acknowledges support from the  Lindheimer fund at Northwestern    
University and from the Polish Science Foundation (FNP) through a 2001
Polish Young Scientist Award.
Research supported in part through KBN grants 2P03D00418, 5P03D01120.

\pagebreak

\begin{deluxetable}{lrrrr}
\tablewidth{250pt}
\tablecaption{Number of coalescing double compact objects 
obtained from $7\times 10^{6}$ initial binaries}
\tablehead{ Primary mass & \multicolumn{4}{c}{ Secondary type} }
\startdata

\, [$M_\odot$]        & WD\tablenotemark{a} & NS\tablenotemark{b}  &  
                        SS\tablenotemark{c} & BH\tablenotemark{d} \\ 
$1.3<M<1.7$           & 32956   & 5533    &         &          \\
$1.7<M<2.5$           & 11305   &  4738   & 166     &          \\
$2.5< M$               & 9650   &  2216   &  1186   &  6291    \\

\enddata
\tablenotetext{a}{white dwarf}
\tablenotetext{b}{compact object with mass $1.3<M<1.7$}
\tablenotetext{c}{compact object with mass $1.7<M<2.5$}
\tablenotetext{d}{compact object with mass $2.5< M$}
\label{tab1}
\end{deluxetable}

\begin{figure}
\plotone{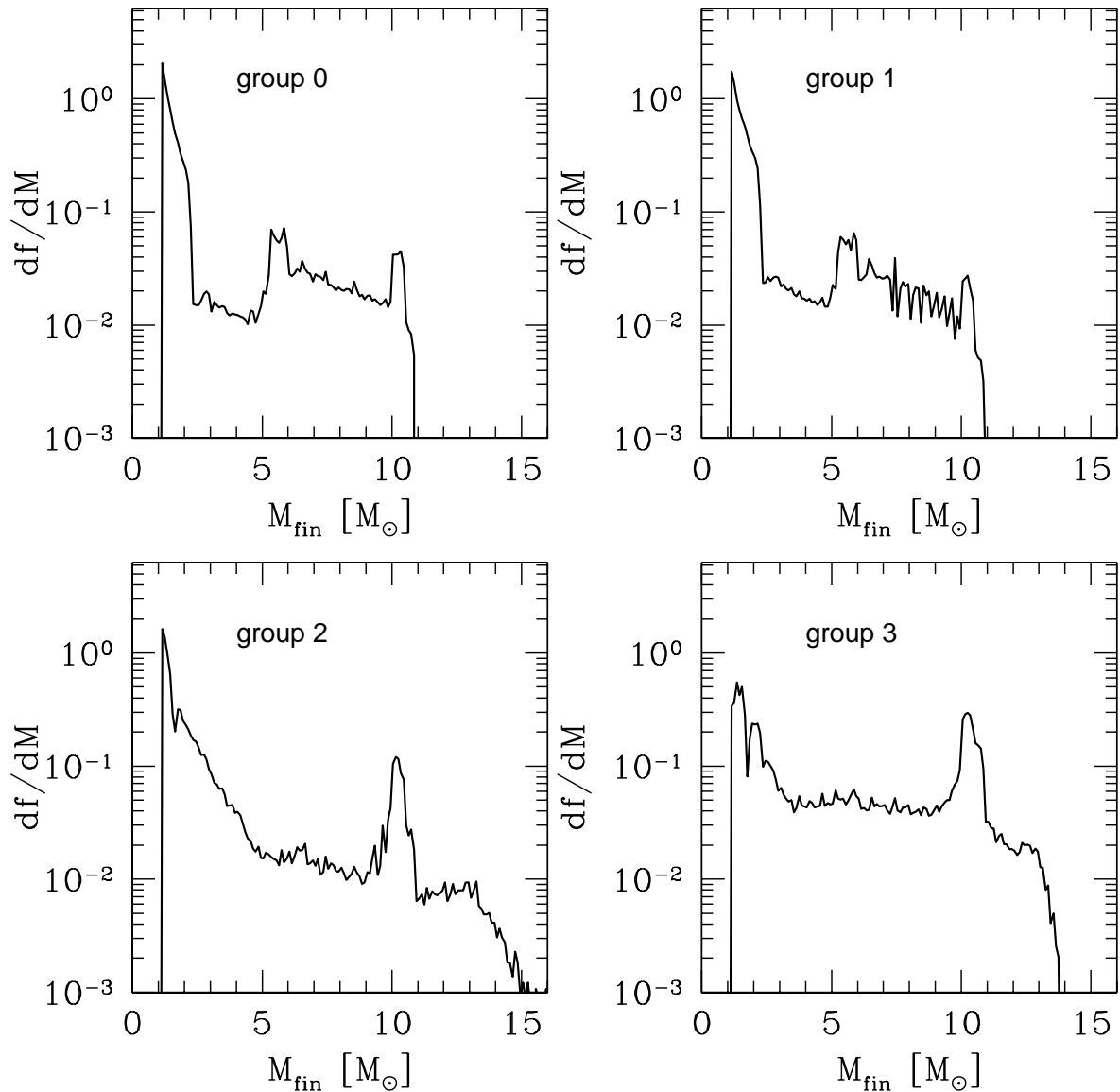}
\caption{The number per unit mass of compact objects (i.e., of neutron
stars, strange stars, or black holes) formed in various ways. 
The top left panel corresponds to the case of single
stellar evolution (group 0), the top right panel represents the 
single compact objects formed in binary evolution (group 1), 
the bottom left panel shows the compact objects in white dwarf 
binaries (group 2), and the the bottom right panel shows those
compact objects whose binary companion is also
a compact object  (group 3).}
\label{fig2}
\end{figure}

\begin{figure}
\plotone{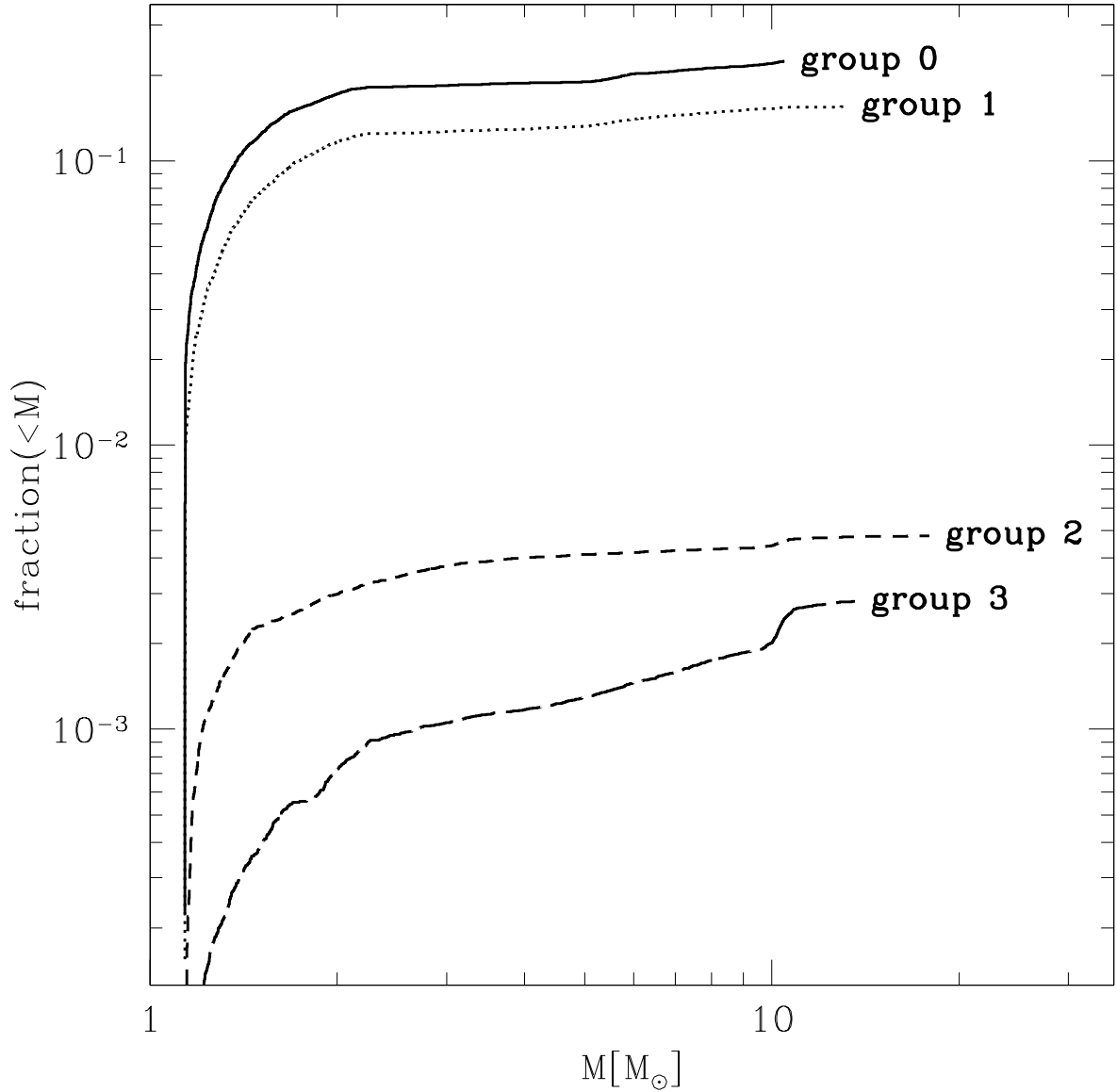}
\caption{The cumulative fraction of compact objects
 corresponding to the 
differential distributions of Figure~\ref{fig1}.
The top curve is for compact objects arising from single stars. The
remaining curves describe the outcome of binary evolution---note
that the most likely fate of a compact object born in a binary is
to be single, and that to end up as a companion to a white dwarf
is more likely than to be in a binary with another compact object.
}
\label{fig2}
\end{figure}

\end{document}